\documentstyle[prl,aps,psfig,epsfig]{revtex}
\begin{document}
\title{\bf 
Transparency Resonances
 and Bound States of the $\delta^\prime$ Junction
}
\author{
 P.L.~Christiansen$^\dag$, A.V.~Zolotaryuk$^{\dag,\ddag}$, 
 V.N.~Ermakov$^{\dag,\ddag}$, and  Y.B.~Gaididei$^{\dag,\ddag}$
}
\address{
$^{\dag}$Section of Mathematical Physics, IMM,
Technical University of Denmark,  DK-2800 Lyngby, Denmark \\
$^{\ddag}$Bogolyubov Institute for Theoretical Physics,
03143 Kyiv, Ukraine
 }
\date{\today}
\wideabs{
\maketitle

\begin{abstract}

Exact positive and negative energy solutions for the 
eigenvalue problem of the 
Schr\"{o}dinger equation in one dimension 
with a $\delta^\prime$ interaction are found
and analyzed. An  infinite series of 
transparency resonance levels in the strength of this 
interaction is shown to exist.
This result is against the actual belief that the
$\delta^\prime$ potential
acts as a totally reflecting wall. A finite
number of bound states is obtained, contrary to the previous 
result on the existence of only one bound state.
A new effect of a {\it negative stepwise} drop in the electron
density across the $\delta^\prime$ junction is observed.
The solutions are also applied to
the propagation of the electromagnetic field in
dielectric  media.

\end{abstract}
}
\bigskip



Point or contact interactions are widely used in various 
applications to quantum physics \cite{do}, producing exactly 
solvable models of complicated physical phenomena
\cite{a-h}.
Intuitively,  these interactions are understood as 
sharply localized potentials, exhibiting a number of 
interesting and intriguing features. 
Applications of these models to solid-state physics
are of particular interest nowadays, mainly
because of the rapid progress
in fabricating nanoscale quantum devices. 
Other applications arise,
for instance, in dielectric media where electromagnetic 
waves scatter at boundaries or thin layers  \cite{ma}.  
In this Letter, we report on a resonance transparency and 
a bound states structure  
of the  derivative of Dirac's delta function potential.
This so-called  $\delta^\prime$ interaction  has  
attracted considerable attention because of a series of 
controversial results published in both 
physics- and mathematics-oriented literature 
\cite{gh,zh,gr,pa,cnp,cs,ehz}, 
 many of which are  not yet fully understood. 
In one dimension, and in the limit that 
 neglects the interactions between electrons,
the $\delta'$ interaction 
 is formally given by the stationary Schr\"{o}dinger
equation 
\begin{equation}
-\psi''(x) + \sigma^2 \delta^\prime(x)\psi(x) 
=p^2\psi(x),
\label{1}
\end{equation}
where the prime denotes the differentiation over 
 the spatial coordinate $x$, $\psi(x)$ is the wave function
for a  particle of mass $m$ (the units in which $\hbar^2/2m =1$
 are used), $\sigma$  the dimensionless strength 
interaction parameter, and  $ p^2 $
(positive, zero, or negative) energy.

It has been proposed ealier \cite{gh} that the
spectral problem of the Schr\"{o}dinger equation
with a point interaction at $x=0$
 can be analyzed in terms of 
the equation $({\cal L}_0\psi)(x) =p^2 \psi(x)$,
${\cal L}_0 \equiv -d^2/dx^2$,  for $x \neq 0$,
with appropriate boundary conditions at $x=0$.
As a result, a number of studies  
 were devoted to finding  correct
 boundary conditions at $x=0$,
where the derivative $\delta'(x)$ was located.
The most striking feature of these results is that
 in some cases the boundary conditions
appear to be irrelevant to  explicit solutions 
and one of these cases was analyzed in Ref.~\cite{pa}.

Since $\delta^\prime (x)$ is a generalized function,
it can be realized as a limit of regularized potentials,
which we denote by $V_l(x)$, with   
 $l$ being a regularization parameter,
that converges to $\delta^\prime(x)$ as $l \rightarrow 0$
in the sense of distributions \cite{gs}.
Then, solving the regularized version of Eq.~(\ref{1}),
\begin{equation}
({\cal L}_l\psi_l)(x)  =p^2\psi_l(x),~~
{\cal L}_l \equiv -d^2/dx^2 +  \sigma^2 V_l(x), 
\label{2}
\end{equation}
we analyze the behavior of the wave function $\psi_l(x)$
 in the limit $l \rightarrow 0$. In this way, 
the subtleties of the boundary conditions at $x=0$
for the operator ${\cal L}_0$ can be avoided. Instead, a rigorous 
mathematical meaning to the limit of the sequence 
 $\{ {\cal L}_l \}$ in the Hilbert space
$L_2(-\infty, \infty)$ must be given. To this end, 
 the notion of the graph limit introduced 
ealier by Glimm and Jaffe \cite{gj} to prove the 
self-adjointness of the Hamiltonian for the relativistic
Yukawa interaction in one spatial dimension, can be used. 
Thus, let $\{ \psi_l \}$ be 
a sequence of vectors such that there exist
$\psi=\lim_{l \rightarrow 0} \psi_l$ and 
$\theta = \lim_{l \rightarrow 0} ({\cal L}_l\psi_l)$,
forming a pair of limit vectors $\{ \psi, \theta \}$.
Since the domain of each ${\cal L}_l$ is dense in 
$L_2(-\infty, \infty)$, the limiting operator ${\cal L}$
 called the {\it graph limit} of $\{ {\cal L}_l \}$ 
is defined by putting $\theta = {\cal L}\psi$. 

In our special case, it is convenient to construct 
a sequence of regularizing functions  $\{ V_l(x) \}$ 
using  step
approximations to the $\delta^\prime(x)$ function.
 First,  based on physical intuition,
we are able to decide how close is the $\delta^\prime$
potential to the actual (regular) potential.
Second,  our procedure demonstrates how  to realize 
explicitly the wave function discontinuity   at $x=0$, 
the problem of recent interest \cite{cs,ehz}. Third,  
 step-like potentials can easily be 
manufactured using, e.g., thin layers of different 
types of semiconductors. More specifically, 
the $\delta^\prime$ interaction may be used as a limiting case
of a device in which a small region of large repulsive potential
is followed by a small region of large attractive potential,
which may be called a {\it barrier-well junction}.  
The similar experimental situation can be arranged in 
dielectric media \cite{ma}. 
Therefore, the results reported here are applicable to
larger classes of wave equations than the 
Schr\"{o}dinger equation.

It is instructive to consider two types of
 regularizing sequences  $\{ V_l(x) \}$, which are
expected to result in different physics in the limit
$l \rightarrow 0$. The key point is to impose  
 different constraints 
 on the behavior of $\{ V_l(x) \}$ 
in the vicinity of $x=0$. For the first type,   
each  function [denoted by $ V^0_l(x)$]
 is supposed to be  identically 
 zero in the vicinity of $x=0$, putting, e.g.,  
 $ V^0_l(x) =[ \delta(x+l) -\delta(x-l)] / (2l)$ \cite{pa},
 or appropriate step functions, the support of which does
not contain the point $x=0$.
Then for each $l>0$, we have an {\it oscillatory}
solution in the vicinity of $x=0$, which collapses
to a {\it continuous} function
 with a node at $x=0$ when $l \rightarrow 0$,
so that an incident current is {\it totally} reflected
at this point \cite{pa}.
Note that for each $l >0$, the energy flow is {\it partially}
transmitted. For the second type of regularizing functions 
 $ V_l^d(x) $,  an opposite constraint at $x=0$,
namely  the existence of a discontinuity at this 
point that goes to infinity as $l \rightarrow 0$, is imposed.
As an example, the step function $V_l^d(x)$ shown in 
Fig.~\ref{fig1} that splits the $x$ axis 
into four regions:
$V^d_l(x) = 0$ ($ |x| > l >0$), 
$l^{-2}$ ($ -l < x < 0$), and  $-l^{-2}$ ($0 < x < l$),
can be chosen.  With this potential,  at the $x=0$ point,  two
different types of a solution are connected, namely tunneling
and oscillatory ones. Therefore for each finite $l$, the wave function
$\psi_l(x)$ has no node at $x=0$,
 and one could expect that
at least for some system parameter values, the limit 
 of continuous functions $\psi_l(x)$ will have a finite
discontinuity at this point. This would immediately implies
 a nonzero current flow across the $\delta^\prime$ junction
 as shown below.

The current (energy flow) 
transmission coefficient $T$ is calculated from
the scattering solution $\psi_l(x)$ 
of Eq.~(\ref{2}) with $p^2>0$
 in a standard way. For the potential depicted in Fig.~\ref{fig1},
it appears to depend on the two dimensionless parameters
$\eta \equiv lp$ and $\sigma$:
\begin{eqnarray}
T & = & 4 \left\{ 2  
+ \left[ \cos(\alpha) \cosh(\beta) - (\alpha/ \beta)
\sin(\alpha) \sinh(\beta) \right]^2 \right. \nonumber \\
&+&   \left[ \cos(\alpha) \cosh(\beta) +  (\beta / \alpha) 
\sin(\alpha) \sinh(\beta) \right]^2 \nonumber \\ 
&+&  \eta^2 \left[ \alpha^{-1} \sin(\alpha) \cosh(\beta) 
+ \beta^{-1} \cos(\alpha) \sinh(\beta)  \right]^2 \nonumber \\ 
&+& \left. \eta^{-2} \left[ \alpha\sin(\alpha) \cosh(\beta) -
 \beta\cos(\alpha) \sinh(\beta) \right]^2 \right\}^{-1}, 
\label{3} 
\end{eqnarray}
where 
$ \alpha = \sqrt{\sigma^2 + \eta^2}$ and  
$\beta = \sqrt{\sigma^2 - \eta^2}$.
Note that the transmission across this junction 
does not depend on the sign 
in front of $\sigma^2$ in Eq.~(\ref{1}), i.e., it is the same
both from the left and from the right.
As shown in Fig.~\ref{fig2}, the transmission over the barrier 
with energies $\eta$ close to the barrier suffers big variations
while the parameter $\sigma$ changes.
 

Let us consider the limit $\eta \rightarrow 0$ that corresponds
to the limit $l \rightarrow 0$. In this case, 
 $\alpha, \beta \rightarrow \sigma$, while
 $T \rightarrow 0$, except for those values
of $\sigma$ that satisfy the equation
\begin{equation}
\tan \sigma = \tanh \sigma .
\label{4}
\end{equation}
This equation, except for the trivial solution 
$\sigma_0 =0$ admits a discrete countable set  of
``resonance'' levels $\sigma_n,~n=1, 2, \ldots ,$ at which 
the $\delta^\prime$ junction becomes   partially {\it transparent},
in contrast to the general belief that the $\delta^\prime$ 
interaction acts as a reflecting wall \cite{pa}.
Indeed, in the limit  $\eta 
\rightarrow 0$, from Eq.~(\ref{3}) one obtains
the limiting values of the transmission coefficient $T$:
\begin{equation}
T_n \equiv \lim_{\eta \rightarrow 0} 
T(\eta, \sigma_n)  
=1-\tanh^4\sigma_n ,~n=1, 2, \ldots .
\label{5}
\end{equation}
One can see from the last equation that 
there is no dependence of $T$ on the dimensionless 
energy parameter $\eta$.
Its behavior is illustrated in Fig.~\ref{fig3} as a function of 
the amplitude parameter $\sigma$ for different values of
the dimensionless parameter $\eta$. It is clearly shown
that the continuous curve $T=T(\sigma)$ as a function of 
$\sigma$ tends to isolated, descreasing 
(in $\sigma$) peaks as $\eta \rightarrow 0$ (see Fig.~\ref{fig3}).


One can easily find the left and the right values of the 
limiting wave function
 $\psi(x)$ as $ x \rightarrow \pm 0$.
Let us denote these values of the wave function
$\psi(x)$ from the barrier and the well sides 
as
$ \psi_b  \equiv  \lim_{l \rightarrow 0} \psi_l(-l)$
 and 
$ \psi_w \equiv \lim_{l \rightarrow 0} \psi_l(l)$,
 respectively.

Using the scattering solution of 
Eq.~(\ref{2}) with $p^2 >0$,  one finds the ratio
\begin{equation}
 \vert \psi_b / \psi_w \vert^2 =  
 (1- \tanh^2 \sigma_n)/ (1+ \tanh^2 \sigma_n) < 1  ,
\label{6}
\end{equation}
which shows that the limiting wave function $\psi(x)$ has a 
{ \it finite nonzero jump} at $x=0$ if $\sigma = \sigma_n$. 
The last inequality means that the probability 
to find an electron nearby the junction is {\it higher}
from the side of the well than that of the barrier. 
In other words, at the resonant levels
$\{ \sigma_n \}_{n=1}^\infty$,
a {\it negative } drop in the electron density
occurs, while passing across 
the junction from the barrier to the well side.

In order to find a discrete spectrum of bound states,
we solve Eq.~(\ref{2}) with negative energy
 $p^2 < 0 $, putting $\zeta \equiv l 
\sqrt{ - p^2}$. As a result, one finds the equation on the 
discrete spectrum $ \zeta_n,~n=0, 1, \ldots : $ 
\begin{eqnarray}
&& (\zeta^2/\alpha\beta)  \tan(\alpha) \tanh(\beta) 
+  (\zeta / 2) \left( \alpha^{-1} \tan\alpha   +
 \beta^{-1} \tanh\beta \right) \nonumber \\
& & + ~ 1 -
( \alpha\tan\alpha - \beta\tanh\beta)/ 2\zeta =0 ,
\label{7}
\end{eqnarray}
where $\alpha =\sqrt{\sigma^2-\zeta^2}$ and  $\beta 
=\sqrt{\sigma^2 +\zeta^2}~$. The
solutions of Eq.~(\ref{7}) with respect to $\zeta$ 
 depend on $\sigma$ and they form a {\it finite discrete}
spectrum $\{ \zeta_n \}_{n=0}^N$ depicted in 
Fig.~\ref{fig4}. The number of bound states $N=N(\sigma)$ 
is determined by the  inequalities $\sigma_N < \sigma
 < \sigma_{N+1}$.
In the vicinity of small $\zeta$, Eq.~(\ref{7}) is 
asymptotically reduced to 
\begin{equation} 
 \alpha \tan\alpha -\beta \tanh\beta =2\zeta ,
\label{7a}
\end{equation}
 from which one finds that in the limit 
$\zeta \rightarrow 0$ ($l \rightarrow 0$), the $n$th eigenvalue 
$\zeta_n(\sigma)$, $n=0, 1, \dots , $
begins at $\sigma=\sigma_n$, i.e., $\zeta_n(\sigma_n)=0$
as shown in Fig.~\ref{fig4}. Note that each eigenfunction
$\psi_n(x)$ that corresponds to the $n$th eigenvalue
with $\sigma = \sigma_n$  has at $x=0$ the same discontinuity jump  
 as the scattering wave function $\psi(x)$ for the same value
 $\sigma_n$.


In quantum mechanics applications, the dimensionless 
quantity $\sigma$ that scales the interaction strength 
is a {\it system} parameter and therefore cannot be 
controlled during electron transport across the 
$\delta^\prime$ junction.  However, 
in integrated optics \cite{ma},  this parameter can appear as 
a frequency of the electromagnetic field. 
 Moreover, 
the results of the spectral analysis obtained above
 for the $\delta^\prime$ interaction can be applied there
directly. Thus, let us consider the Maxwell equations in 
a medium with one-dimensional (say, along the $x$ axis as
shown in Fig.~\ref{fig5})  inhomogeneity of
 dielectric permeability.
In the Cartesian coordinate system, for TE waves
($E_x = H_y =E_z=0$) with frequency $\omega$, the $E_y$ 
component satisfies the equation
\begin{equation}
{ \partial^2 E_y \over \partial x^2}
+ { \partial^2 E_y \over \partial z^2} 
+k^2 \varepsilon(x)E_y=0 ,
\label{8}
\end{equation}
where $k =\omega\sqrt{\varepsilon_0 \mu}$ is 
the wave number with 
$\varepsilon_0$ being the dielectric permeability of
vacuum and $\mu$ the magnetic permeability, and 
$\varepsilon(x)$ is a dimensionless profile of 
dielectric permeability along the $x$ axis.
Assuming next a wave propagation along the $z$ axis, i.e., 
putting $E_y(x,z)=E(x)\exp(iqz)$ where $q$ is a wave vector, 
one obtains the eigenvalue problem:
\begin{equation}
E''(x) +k^2 \varepsilon(x) E(x) = q^2 E (x),
\label{9}
\end{equation}
with two spectral parameters: wave numbers  $k$ and $q$.

We consider the $\varepsilon(x)$ profile of the form
shown in  Fig.~\ref{fig5}, where
$\varepsilon(x) =\varepsilon_b$ if $|x|> a$, and 
$\varepsilon(x) =\varepsilon_b + \varepsilon_m $ if 
$0 < x \le a $ and 
$\varepsilon(x) =\varepsilon_b -  \varepsilon_m $ if 
$-a \le x < 0$.
Using the scaling transformation $x \rightarrow a x/l$
and  $E(x) \rightarrow E(ax/l) \equiv \psi_l(x)$, 
Eq.~(\ref{9}) can be  
reduced to Eqs.~(\ref{2}), in which 
\begin{equation}
\sigma^2 = k^2a^2 \varepsilon_m~~\mbox{and}~~  
p^2 = (a/l)^2 ( k^2\varepsilon_b -q^2).
\label{10}
\end{equation} 
 Then in Eqs.~(\ref{3}) and (\ref{7}), we have to put
$\eta = a \sqrt{k^2 \varepsilon_b -q^2}$
and $\zeta = a \sqrt{q^2 -k^2\varepsilon_b}~$,
respectively. Therefore, the {\it cut-off} frequencies
$\omega$ (or $k$) 
that cut the discrete spectrum of waveguide modes from 
the continuous spectrum  ($\eta=\zeta=0$) lie on the line 
$q=k \sqrt{\varepsilon_b}$,
as shown by line 1 in Fig.~\ref{fig6}. More precisely, they
 are given by the points $(k_n,q_n^0)$ 
on the $(k,q)$ plane, where 
$k_n =\sigma_n/ (a \sqrt{\varepsilon_m}) $ and 
$q_n^0=(\sigma_n/a) \sqrt{\varepsilon_b/\varepsilon_m}~$,
with $\sigma_n$'s, $n=1,~2, \ldots ,$ being 
positive roots of Eq.~(\ref{4}). Moreover, according to
 Eq.~(\ref{7a}),
one can conclude that the cut-off  and 
the transparency frequencies  coincide. 
In other words,  the junction of the 
form shown in Fig.~\ref{fig5} is partially transparent
in the transverse direction at the cut-off frequencies
of the waveguide regime that occurs in the longitudinal direction.
These results are illustrated by Fig.~\ref{fig6}.
Using Eqs.~(\ref{10}),  one finds that 
the waveguide regime is given by the 
set of curves $q_n(k)= 
\sqrt{k^2\varepsilon_b +\zeta_n^2(ka\sqrt{\varepsilon_m})
/a^2}$ which are depicted above line 1 in Fig.~\ref{fig6}.
The equation $q=k\sqrt{\varepsilon_b -\varepsilon_m}$
splits (see line 2 in Fig.~\ref{fig6}) 
the whole sector of 
continuum spectrum into two sectors I and II that
correspond to $ \eta < \sigma$ and $\eta > \sigma $, respectively.
In these sectors, the transparency properties given by
the transmission coefficient $T(\eta,\sigma)$ are demonstrated
 in Fig.~\ref{fig2}.


Thus, using the sequence of physically motivated 
regularizing functions shown in Fig.~\ref{fig1}, 
 an infinite, countable set of resonance values for the
interaction strength parameter $\sigma$, at which a  
transparent regime of current flow occurs, 
has been found explicitly
 for the $\delta^\prime$ interaction 
described by the Schr\"{o}dinger equation (\ref{1}). All these 
resonance levels, $\{ \sigma_n \}_{n=1}^\infty$,
appear to be positive roots of the simple equation (\ref{4}).
This result contradicts Patil's result \cite{pa},
according to which the transparency is identically zero.
 The reason of this controversy 
emerges from the fact that the approximating sequence 
 $\{ V_l^0(x)\}$ constructed from 
the double of Dirac's $\delta$ functions \cite{pa}, 
necessary results in a {\it continuous} wave function with a 
{\it node} at $x=0$ as $l \rightarrow 0$, blocking a 
current across this point. A controversy also arises
concerning the bound states structure. Instead of 
{\it one} bound state as claimed previously \cite{pa},
a finite number of such states is found, increasing with
$\sigma$. The profiles
 of the bound states are shown to have also finite jumps 
at $x=0$. 

Finally, note that the graph
limit ${\cal L}$ cannot be represented 
as the sum of ${\cal L}_0$ 
and a well-defined object $\delta^\prime(x)\psi(x)$. 
Since $\psi(x)$ is a discontinuous function at $x=0$,
this product is meaningless, but in the regularization 
process, the {\it infinite} 
 renormalization of $V_l^d(x)\psi_l(x)$ occurs, so that
 divergent terms in the limiting operator ${\cal L}$ 
are totally cancelled. 

The authors acknowledge partial support
from the European Union under the INTAS
Grant No.~97-0368 and the RTN Project 
LOCNET HPRN-CT-1999-00163.
We thank S{\o}ren Christiansen, Jens Juul Rasmussen,
Alwyn C. Scott, and Yaroslav Zolotaryuk 
 for stimulating and helpful discussions.

\newpage
\section{Figure captions}

\begin{figure}[htb]
\caption{
The piecewise constant approximation of the $\delta^\prime(x)$ 
function, where $l$ is the width (regularization) parameter.
}
\label{fig1}
\end{figure}

\begin{figure}[htb]
\caption{
Dependence of the 
transmission coefficient $T$ on $\eta$ and $\sigma$:
(a) Three-dimensional surface $T(\eta,\sigma)$, where
 the white curve shows the values of $T$ on the line
$\eta=\sigma$. (b) Section of the surface $T(\eta,\sigma)$
at $\eta=15$. The inset 
demonstrates big variations of the coefficient 
$T$ nearby $\eta=\sigma$.  
}
\label{fig2}
\end{figure}

\begin{figure}[htb]
\caption{
Logarithm of the transmission coefficient $T$ against $\sigma$ at 
 $\eta=0.0005$ (solid curve), 0.05 
(dashed curve), and 0.5  (dashed-dotted curve).
}
\label{fig3}
\end{figure}

%
\begin{figure}[htb]
\caption{ 
Five bound states $\zeta_n,~n=0, 1, 2, 3, 4,$ as functions
of $\sigma$ depicted up to $\sigma =15$.
}
\label{fig4}
\end{figure}

%
%
\begin{figure}[htb]
\caption{ 
Profile of the dimensionless step  dielectric permeability 
$\varepsilon(x)$ along the $x$ axis; 
$2a$ is the width of the junction layer,
$\varepsilon_b$  a  background dielectric permeability, 
and $\varepsilon_m$ the amplitude of permeability variation
across the layer.
}
\label{fig5}
\end{figure}

%
\begin{figure}[htb]
\caption{
Discrete  and continuous 
plane spectra of electromagnetic 
wave transmission across the layer structure
shown in Fig.~\ref{fig5}. 
The cut-off frequencies $(k_n, q_n^0)$
lie on the line $q= k\sqrt{\varepsilon_b}$ labeled by 1.
Above this line, the first four waveguide modes
are shown by  curves $q_n(k)$, 
$n=0, 1, 2, 3.$ Below, in the continuous spectrum sector,
the line $q=k\sqrt{\varepsilon_b-\varepsilon_m}$ labeled by 2 
represents the scattering regime nearby the minimum
of dielectric permeability,  $\varepsilon =
\varepsilon_b - \varepsilon_m$ .
}
\label{fig6}
\end{figure}

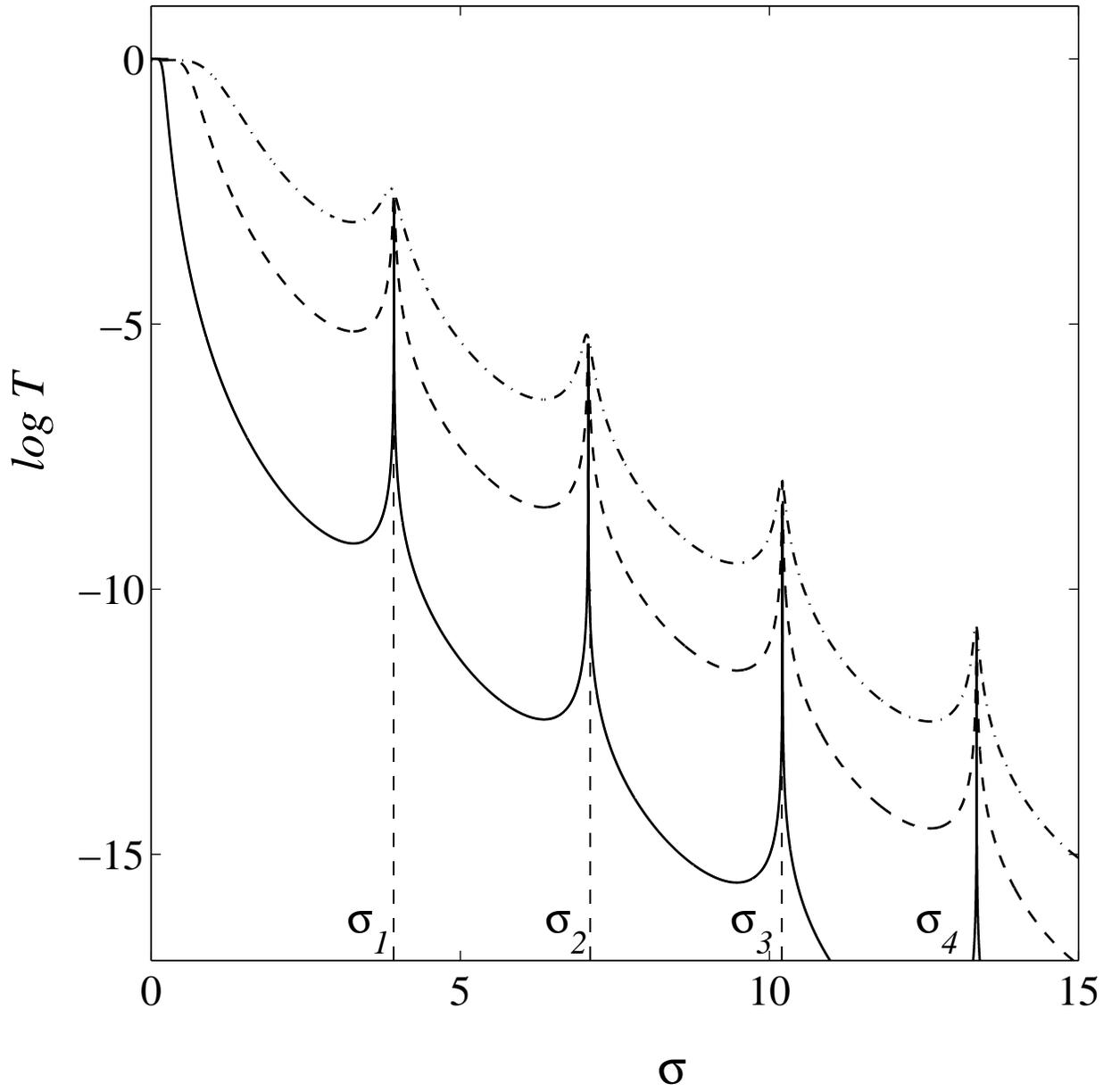
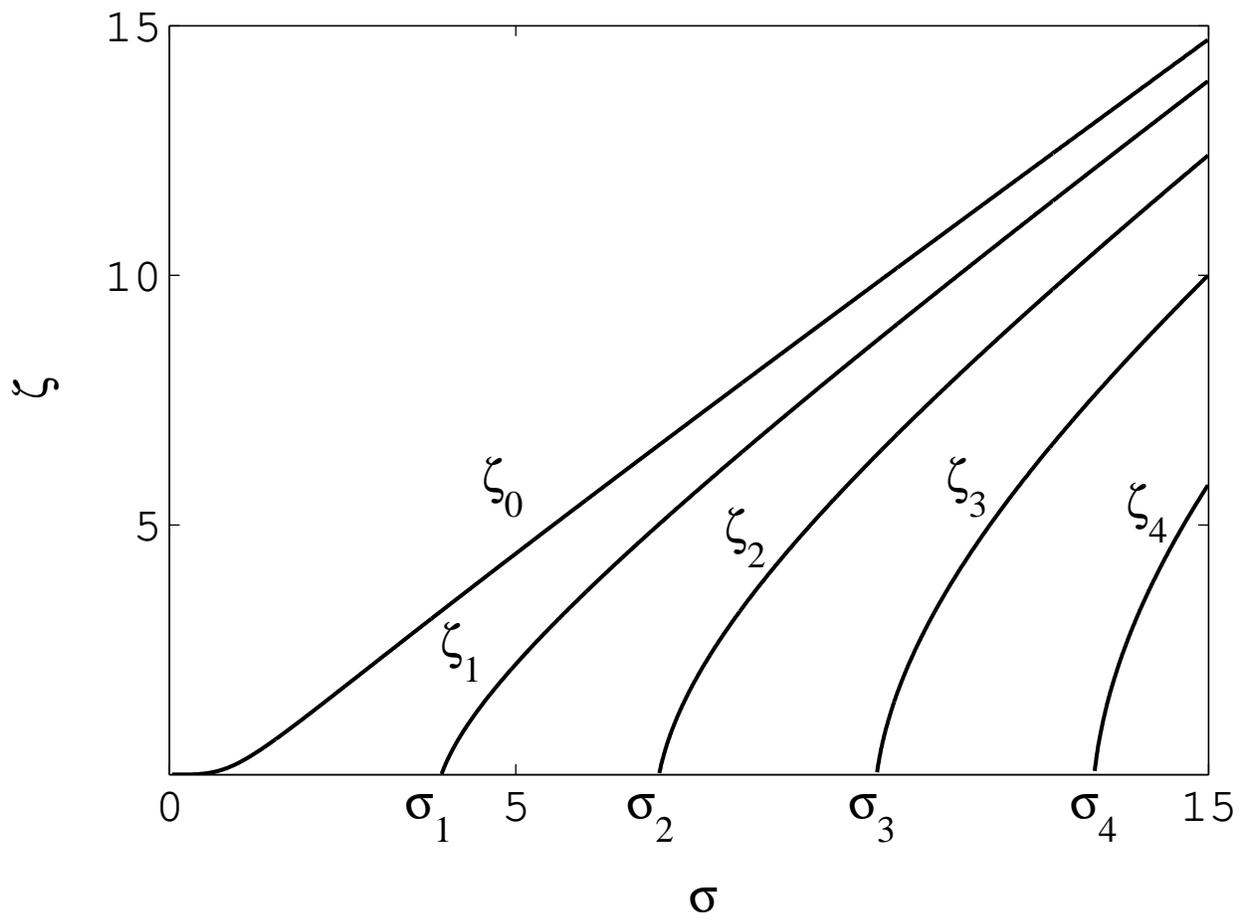
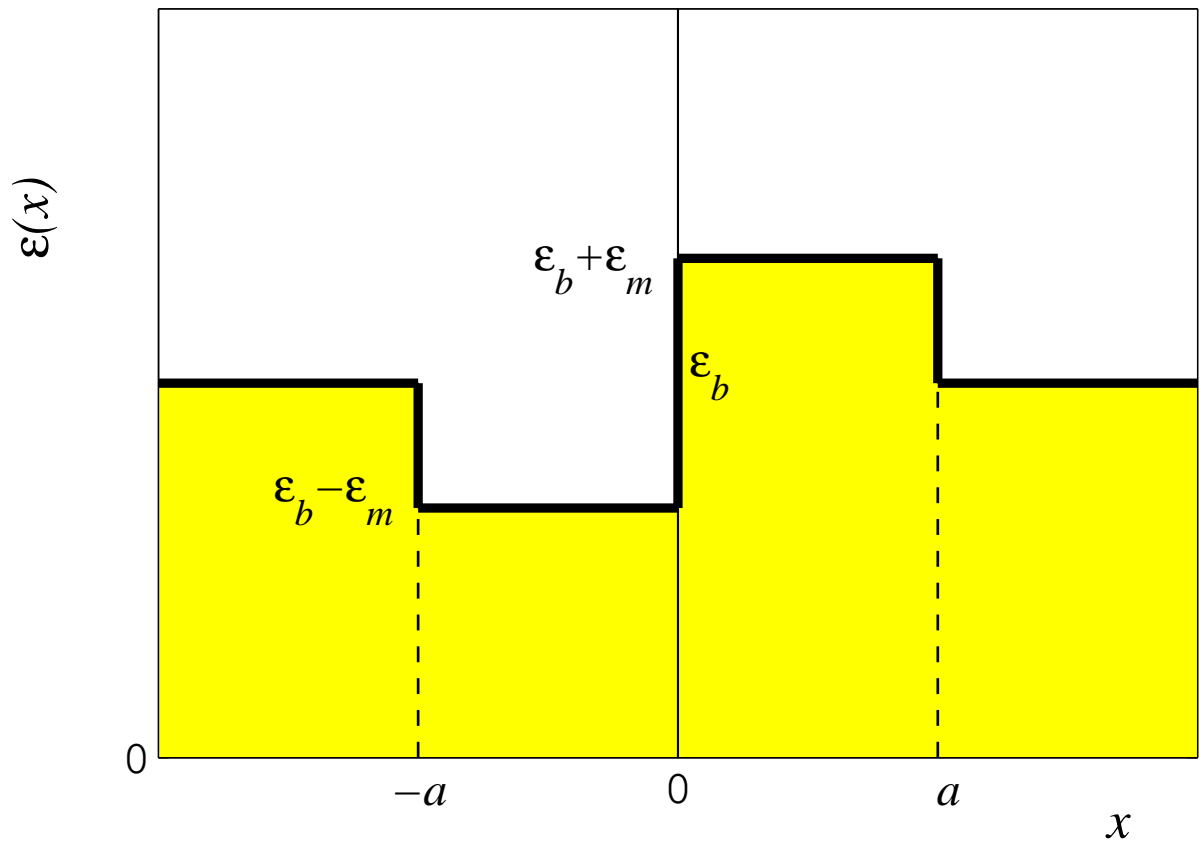
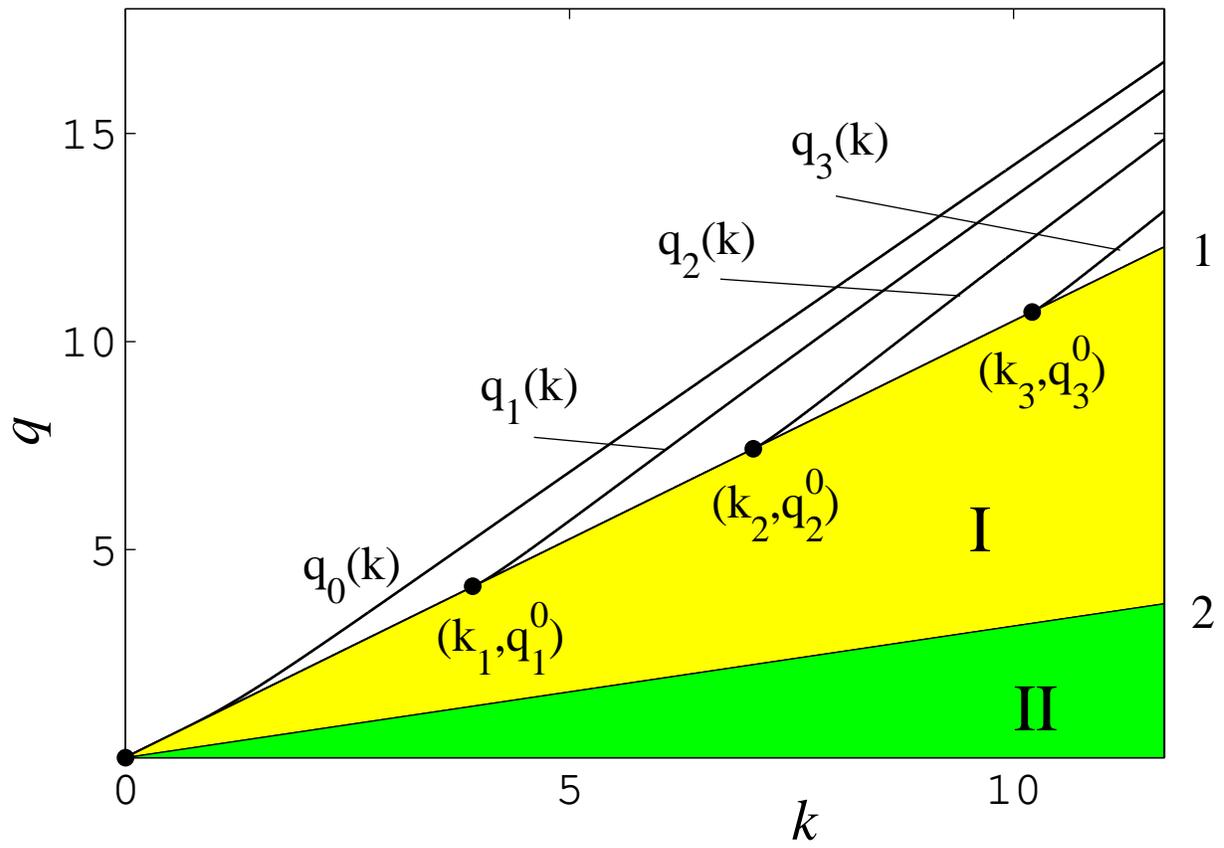
\end{document}